\documentstyle[12pt]{article}
\begin{document}
\begin{titlepage}

\centerline{\large \bf Evolution of chiral-odd spin-independent}
\centerline{\large \bf fracture functions in Quantum Cromodynamics.}
\vspace{10mm}

\centerline{\bf A.V. Belitsky and E.A. Kuraev}
\vspace{10mm}

\centerline{\it Bogoliubov Laboratory of Theoretical Physics}
\vspace{3mm}
\centerline{\it Joint Institute for Nuclear Research}
\vspace{3mm}
\centerline{\it 141980, Dubna, Russia}
\vspace{20mm}

\centerline{\bf Abstract}

\hspace{0.5cm}

We construct the evolution equations for the twist-3 chiral-odd
spin independent fracture functions in QCD. The Gribov-Lipatov 
reciprocity relation is fulfilled at the one-loop level for the 
quasi-partonic two-particle cut vertices only. It is found that 
the rang of the anomalous dimensions matrix is infinite for any 
given moment of the three-parton fracture function as distinguished
from the case of DIS-distributions where the rang of the matrix was
finite and increases with the number of the moment. In the multicolour 
limit $N_c \to \infty$ the evolution equation for the quark-gluon-quark 
correlation function decouples from another equation in the system and 
becomes homogeneous provided we discard the quark mass effects. This fact 
provides an opportunity to find its analytic solution explicitly in nonlocal 
form similarly to the DIS. 
\end{titlepage}


\section{Introduction.}

The deep inelastic scattering (DIS) of leptons on the hadron target
is the most effective experimental tool for studying the dynamics
of hadron reaction on the parton level which has a firm basis
in the quantum field theory provided by light-cone Operator Product
Expansion (OPE) \cite{wil69}. The latter gives a strict theoretical 
ground for separation
of two different scales of the underlying process: nonperturbative
information is concentrated in the tower of local composite
operators while the coefficient function which characterizes the
hard interaction process of constituents can be dealt 
perturbatively. It makes possible the study of logarithmic
violation of Bjorken scaling as well as the power suppressed
contributions (higher twists) responsible for many subtle 
phenomena in a polarized scattering. However, there exists an
equivalent approach to the analysis of the corresponding 
quantities which is based on factorization theorems \cite{mue89} 
and the evolution equations \cite{lip74,alt77}. In spite of the fact that 
the latter 
approach has some difficulties as compared to the former in the
study of higher twists, like the loose of the explicit gauge and Lorentz 
invariance of calculations and also the presence of the 
overcomplete set of correlation functions, it has the important 
advantage as being the most close to the intuitive physical 
picture and the similarity to the parton model. 
There is another advantage of the latter approach for studying of
the higher twist effects from the point of 
view of experimental capabilities since the OPE provides us for 
the moments of the structure functions and in order to extract the
former one needs to measure the latter in the whole region of the 
momentum fraction very accurately. Obviously, it is quite difficult task
even for next generation colliders. While, with a set of evolution
equations at hand, one can find, in principle, the $Q^2$-dependence 
of the cross section in
question by putting the experimental cuts on the region of the 
attained momentum fractions. This approach can also be used in the
situations when the OPE is no longer valid. These are the inclusive 
production of the hadron in the $e^+e^-$-annihilation, 
semi-inclusive deep inelastic scattering, Drell-Yan lepton
production {\it et ctr}.

There is continuous interest in the inclusive production of 
hadron in hard reaction. This process involves a quark fragmentation
function to describe the hadron production from the underlying 
hard parton scattering. But it differs considerably from the
DIS as the short distance expansion is no longer relevant although
the given process goes near the light cone. The theoretical basis 
for strict analyses of the above phenomena is provided by the
generalization of the OPE to the time-like region in terms of 
Mueller's $\zeta$-space cut vertices \cite{muel78,muel81}, which 
moments are essentially nonlocal in the coordinate space. Again this 
approach has all attractive features of the OPE as it provides 
a consistent framework to account for the higher twist effects 
\cite{bal91} as well as it allows to sum up the UV logarithms
\cite{cfp80} by using the powerful methods of the renormalization 
group.

The semi-inclusive hadron production from a quark fragmentation
is described in QCD by the specific nonperturbative correlation
functions of quark and gluon field operators over the hadron
states which can be identified with $\zeta$-space cut vertices. While 
the behaviour of the latter with respect to the
fraction of the parton momentum carried by the hadron
is determined by the nonperturbative strong interaction dynamics, 
the large $Q^2$-scale dependence is governed by perturbation 
theory only.

In this paper we begin our study of the evolution of the chiral-odd
and chiral-even twist-3 fragmentation functions. We begin with 
nonpolarized twist-3 fracture functions. This is the simplest 
case with respect to the number of correlation functions involved in 
mixing under renormalization group evolution\footnote{For completeness 
we should note note that the $Q^2$-evolution
of the chiral-odd nonpolarized and polarized distributions was 
investigated in ref. \cite{koi95} in the framework of the OPE by studying
the mixing of composite local operators under renormalization, 
while the coordinate space evolution equations 
in multicolour QCD were constructed
and solved in ref. \cite{bbkt96}}. From the phenomenological point of 
view they appear, 
for example, in the cross section for semi-inclusive hadron production in the
process of measuring the nucleon's transversity distribution $h_1(x)$
through deep inelastic scattering \cite{ji94}:
\begin{eqnarray}
&&\hspace{-0.7cm}\frac{d^4\Delta\sigma}{dxdyd(1/\zeta )d\phi}\\
&&\hspace{-0.7cm}=\frac{4\alpha^2_{em}}{Q^2}
\left[
\cos \chi \left( 1-\frac{y}{2} \right)G_1(x,\zeta)
+ \cos\phi\sin\chi \sqrt{(\kappa-1)(1-y)}
\left( G_T (x,\zeta) - G_1(x,\zeta)\left( 1-\frac{y}{2} \right) \right)
\right]\!\! ,\nonumber
\end{eqnarray}
here $\kappa = 1 + 4 x^2M^2/Q^2$, $y=1-E'/E$ and 
the cross section is expressed in a frame where the lepton beam
with energy $E$ defines the $z$-axis and the $x-z$-plane contains the 
nucleon polarization vector, which has the polar angle $\chi$ and the 
scattered electron $E'$ has the polar angles $\theta,\hspace{0.1cm}\phi$.
The functions $G_1$ and $G_T$ are expressed in terms of the product of the
distribution and fragmentation functions in the following way
\begin{eqnarray}
G_1(x,\zeta)&=&\frac{1}{2}\sum_{i}Q_i^2 g_1^i(x){\cal F}^i(\zeta),\\
G_T(x,\zeta)&=&\frac{1}{2}\sum_{i}Q_i^2
\left[ g_T^i(x){\cal F}^i(\zeta)
+\frac{1}{x}h_1^i(x){\cal I}^i(\zeta) \right].
\end{eqnarray}
All of them have the expressions in QCD in terms of the 
light-cone Fourier transformation of correlation functions
of fundamental quark and gluon fields over specific 
hadron states \cite{ji94,jaf91}
\begin{eqnarray}
S_+g_1(x)&=&\frac{1}{2}\int \frac{d\lambda}{2\pi}e^{i\lambda x}
\langle N |\bar\psi (0)\gamma_+\gamma_5\psi (\lambda n)
|N \rangle, \nonumber\\
S^\perp_\mu g_T(x)&=&\frac{1}{2}\int \frac{d\lambda}{2\pi}e^{i\lambda x}
\langle N |\bar\psi (0)\gamma^\perp_\mu \gamma_5\psi (\lambda n)
|N \rangle, \nonumber\\
S^\perp_\mu h_1(x)&=&\frac{1}{2}\int \frac{d\lambda}{2\pi}e^{i\lambda x}
\langle N |\bar\psi (0)i\sigma^\perp_{\mu +}\gamma_5\psi (\lambda n)
|N \rangle, \nonumber\\
{\cal F}(\zeta)&=&\frac{1}{4\zeta}
\int \frac{d\lambda}{2\pi}
e^{i\lambda\zeta}
\langle 0 |
\gamma_+
\psi (\lambda n) |h,X \rangle
\langle h,X|\bar \psi (0) |0 \rangle .
\end{eqnarray}
and ${\cal I}$ is given by equation (\ref{i}). The summing
over final state hadrons in the definition of the fracture functions 
is implicit throughout the paper.
Note, that the physical regions are different for distribution
and fragmentation functions: $0\leq x\leq 1$ and $1\leq\zeta<\infty$,
respectively.

In the present study we address ourselves to the problem of construction 
the evolution equations for the function ${\cal I}$ which mixes with other 
cut vertices during the renormalization. The outline of the paper is 
the following. In section 2 we give the definitions of the $\zeta$-space cut 
vertices which are closed with respect to the renormalization group 
evolution and discuss the advantages of the light-cone gauge. In the 
third section we show our technique on an example of abelian gauge 
theory: we review the renormalization of the theory in the axial gauge 
and construct corresponding equations. In the section 4 the $Q^2$-evolution 
equations are derived for the QCD case. The final section is devoted to the 
discussion of our results and conclusions. In the appendix we present some 
useful formulae which make the discussion more transparent. 

\section{Definitions.}

It is well known that in order to endow with parton-like interpretation
of field theoretical quantities and to get much deeper insight into 
the corresponding perturbative calculations it is necessary to use 
to ghost-free gauges. Owing to this fact we chose in what follows the 
light-cone gauge $B_+ =n^\mu B_\mu = 0$ for the boson field and make 
the Sudakov decomposition of the parton four-momentum into transverse 
and longitudinal components
\begin{equation}
k^\mu = z p^\mu + \alpha n^\mu + k_\perp^\mu,
\end{equation}
where $p$ and $n$ are null vectors which pick out two different
directions on the light cone such that $p^2=n^2=0$, $(pn)=1$. 
The advantage of this gauge is that the gluon field operator $B_\rho$ 
is related to the field strength tensor $G_{\rho\sigma}$ via simple 
relation
\begin{equation}
B_{\mu}(\lambda n) =\partial^{-1}_+ G_{+ \mu}(\lambda n)
=\frac{1}{2}\int_{-\infty}^{\infty}dz
\epsilon (\lambda - z)G_{+\mu}(z),\label{G}
\end{equation}
so, that the gauge invariant result can be restored after all
required calculations have been performed. 

For our purposes it is much more suitable to deal with correlation 
functions listed below\footnote{See discussion of this point in 
refs. \cite{lip84,lip85}}
\begin{eqnarray}
&&\hspace{-0.7cm}{\cal I}(\zeta)
=\frac{1}{4}\int \frac{d\lambda}{2\pi}
e^{i\lambda\zeta}\label{i}
\langle 0 |
\psi (\lambda n) |h,X \rangle
\langle h,X|\bar \psi (0) |0 \rangle ,\\
&&\hspace{-0.7cm}{\cal M}(\zeta)
=\frac{1}{4\zeta}\int \frac{d\lambda}{2\pi}
e^{i\lambda\zeta}
\langle 0 |
m\gamma_+
\psi (\lambda n) |h,X \rangle
\langle h,X|\bar \psi (0) |0 \rangle ,\\
&&\hspace{-0.7cm}{\cal Z}_1^{(1)}(\zeta',\zeta)
=\frac{1}{4\zeta}\int \frac{d\lambda}{2\pi} \frac{d\mu}{2\pi}
e^{i\lambda\zeta-i\mu\zeta'}
\langle 0 | {\rm g}\gamma^\perp_\rho \gamma_+\!
\psi (\lambda n) |h,X \rangle
\langle h,X|\bar \psi (0) B^\perp_\rho (\mu n)|0 \rangle ,\\
&&\hspace{-0.7cm}{\cal Z}_{1}^{(2)}(\zeta,\zeta')
=\frac{1}{4\zeta}\int \frac{d\lambda}{2\pi} \frac{d\mu}{2\pi}
e^{i\mu\zeta'-i\lambda\zeta}
\langle 0 |
{\rm g}\gamma_+\gamma_\rho^\perp B^\perp_\rho (\mu n)
\psi (0) |h,X \rangle
\langle h,X|\bar \psi (\lambda n) |0 \rangle ,\\
&&\hspace{-0.7cm}{\cal Z}_2^{(1)}(\zeta,\zeta')
=\frac{1}{4\zeta}\int \frac{d\lambda}{2\pi} \frac{d\mu}{2\pi}
e^{i\mu\zeta'-i\lambda\zeta}
\langle 0 |
\bar \psi (0)
{\rm g}\gamma_+ \gamma_\rho^\perp \psi (\mu n) |h,X \rangle
\langle h,X| B^\perp_\rho (\lambda n) |0 \rangle ,\\
&&\hspace{-0.7cm}{\cal Z}_2^{(2)}(\zeta',\zeta)
=\frac{1}{4\zeta}\int \frac{d\lambda}{2\pi} \frac{d\mu}{2\pi}
e^{i\lambda\zeta-i\mu\zeta'}
\langle 0 | B^\perp_\rho (\lambda n) |h,X \rangle
\langle h,X| \bar \psi (\mu n)
{\rm g}\gamma_\rho^\perp \gamma_+ \psi (0) |0 \rangle .\label{last}
\end{eqnarray} 
The quantities determined by these equations form the closed set
under the renormalization, however, they are not independent 
since there is relation between them due to equation
of motion for the Heisenberg fermion field operator
\begin{equation}
{\cal I}(\zeta)-{\cal M}(\zeta) \label{eqmot}
+\int d\zeta' {\cal Z}_1(\zeta',\zeta)=0.
\end{equation}
Here and in the following discussion we introduce the convention
\begin{eqnarray}
&&{\cal Z}_j (\zeta',\zeta)
=\frac{1}{2}
\left[ {\cal Z}_j^{(1)}(\zeta',\zeta)+
{\cal Z}_j^{(2)}(\zeta,\zeta')
\right].
\end{eqnarray}
While the former two functions ${\cal I}$ and ${\cal M}$ can be made 
explicitly gauge invariant by inserting the $P$-ordered exponential
(which is unity in the gauge we have chosen) between the quark fields, 
the latter can be written in the gauge 
invariant way introducing the following objects:
\begin{eqnarray}
{\cal R}_1(\zeta',\zeta)=\zeta' {\cal Z}_1(\zeta',\zeta),\hspace{0.8cm}
{\cal R}_2(\zeta,\zeta')=\zeta {\cal Z}_2(\zeta ,\zeta').
\end{eqnarray}
Taking into account eq. (\ref{G}) it is easy to verify that they
are indeed expressed in terms of correlators involving the gluon field 
strength tensor. The functions $ {\cal Z}^{(1)}$ and ${\cal Z}^{(2)}$ 
are related by complex conjugation
\begin{eqnarray}
&&\left[{\cal Z}_1^{(1)}(\zeta',\zeta)\right]^*
={\cal Z}_1^{(2)}(\zeta,\zeta'),\hspace{0.5cm}
\left[{\cal Z}_2^{(1)}(\zeta,\zeta')\right]^*
={\cal Z}_2^{(2)}(\zeta',\zeta).
\end{eqnarray}
Their support properties can be found by applying Jaffe's recipe 
\cite{jaf83}. It has been shown that the field operators 
entering the definition of the correlation functions can be 
placed in the arbitrary order on the light cone with appropriate 
sign change according to their statistics. Then taking 
the particular ordering and saturating the correlation function by 
the complete set of the physical states we immediately obtain
(for definiteness, we consider the function ${\cal Z}_1^{(1)}$)
\begin{eqnarray}
{\cal Z}_1^{(1)}(\zeta' ,\zeta)=\!\!\!\!\!&&\frac{1}{4\zeta}\sum_{X,Y}
\delta (\zeta - 1 -\zeta_X)\delta (\zeta' - \zeta_Y)
\langle 0|\psi |h,X \rangle
\langle h,X | \bar\psi |Y \rangle
\langle Y | B^\perp |0 \rangle\nonumber\\
=&&\!\!\!\!\!\frac{1}{4\zeta}\sum_{X,Y}
\delta (\zeta - 1 -\zeta_X)\delta (\zeta - \zeta' - \zeta_Y)
\langle 0|\psi |h,X \rangle
\langle h,X | B^\perp |Y \rangle \label{pos}
\langle Y | \bar\psi |0 \rangle,
\end{eqnarray}
with $\zeta_X, \zeta_Y \geq 0$ and we omit the unessential Dirac 
matrix structure of the vertex. From these equations the 
restrictions emerge on the allowed values of the momentum fractions:
$1\leq \zeta <\infty$, $0\leq \zeta'\leq \zeta$. By analogy one can 
easily derive similar support properties for other functions.

\section{Construction of the evolution equations.}

Due to ultraviolet divergences of the momentum integrals in the
perturbation theory there is logarithmic dependence of the
parton densities on the normalization point. This dependence
is governed by the renormalization group. The evolution equations 
for the leading twist correlation functions determining their $Q^2$ 
dependence can be interpreted in terms of 
the kinetic equilibrium of partons inside a hadron (for distribution 
functions) or hadrons inside a parton (for fragmentation function) 
under the variation of the ultraviolet transverse momentum cut-off 
\cite{lip74}.
However, beyond the leading twist the probabilistic picture
is lost due to quantum mechanical interference and more
general quantities emerge, {\it i.e.} multiparticle parton correlation
functions, whose scale dependence are determined by Faddeev type
evolution equation with pairwise particle interaction 
\cite{lip84,lip85}.

There are two sources of the logarithmic dependence of the
correlation functions. The first is the divergences of the
transverse momentum integrals of the particles interacting
with the vertex and forming the perturbative loop. Another 
source is the divergences due to the virtual radiative 
corrections. In the renormalizable field theory the latter
are factorized into the renormalization constants of the
corresponding Green functions. However, owing to the specific
features of the renormalization in the light-like gauge,
extensively reviewed in the next section, there is mixing
of correlation functions due to the renormalization of field
operators. The latter fact is closely related to the matrix
nature of renormalization constants of the elementary Green
functions in the axial gauge. For example, after renormalization
of the fermionic propagator the matrix structure of the bare
cut vertex could be changed, in general, since the renormalization
matrix acts on the spinor indices of the vertex (see eq. 
(\ref{renorm})).

In the Leading Logarithmic Approximation (LLA) there is a strong 
ordering \cite{grlip72_1,grlip72_2} of transverse particle 
momenta as well as their minus components, so that only the 
particles entering the divergent virtual block $\Sigma,\Gamma,\Pi$ 
or the particles adjacent to the cut vertex can achieve the 
maximum values of $|k_\perp |$ and $\alpha$:
\begin{eqnarray}
&|k_{n \perp}| \ll ... \ll
|k_{2 \perp}| \ll |k_{1 \perp}| \ll \Lambda^2 ,& \nonumber\\
&\alpha_n \ll ... \ll \alpha_2 \ll \alpha_1 ,&
\end{eqnarray}
while the plus components of the parton momenta are of the same 
order of magnitude
\begin{equation}
z_n \sim ... \sim z_2 \sim z_1 
\end{equation}
for $n$-rank ladder type diagram.

The radiative corrections to the bare cut vertex can be calculated
using the conventional Feynman rules with the following 
modifications:
\begin{itemize}

\item All propagators and vertices on the RHS of the cut are hermitian
conjugated to that on the LHS.

\item Every time crossing the cut have the propagator $1/(k^2-m^2+i0)$
replaced by $-2\pi i\delta (k^2-m^2)$.

\item For each propagator crossing the cut there is a $\theta$-function
specifying that the energy flow from the LHS to the RHS is positive.
(In the infinite momentum frame this is the plus component of the
four-momentum.)
\end{itemize}
These statements complete the rules to handle the cut vertices.

\section{Abelian evolution}

In this section we show our machinery on a simple example
of abelian evolution and generalize it afterwards to the Yang-Mills 
theory. We start with overfull set of the cut vertices defined by 
eqs. (\ref{i})-(\ref{last}) and disregard for a moment the relation 
between them. Then eq. (\ref{eqmot}) verifies that the evolution 
equations thus obtained are indeed correct. We have to note that
since the observed particle $h$ is always in the final state some 
cuts of Feynman diagrams are not allowed and, therefore, we could 
not obtain the evolution kernel for the cut vertex taking the 
discontinuity of uncut graph as we are restricted over the limited 
set of the cuts.

\subsection{Renormalization in the light-cone gauge.}

The peculiar feature of the light-cone gauge manifest itself 
in the existence of additional UV divergences of the Feynman graphs
which are absent in the usual isotropic gauges. They appear due to 
particular form of the gluon propagator
\begin{eqnarray}
&&D_{\mu\nu}(k)=\frac{d_{\mu\nu}(k)}{k^2+i0}\nonumber\\
&&d_{\mu\nu}= g_{\mu\nu}-\frac{k_\mu n_\nu + k_\nu n_\mu}{k_+}
\end{eqnarray}
that possesses an additional power of the transverse momentum
$k_\perp$ in the numerator. 
For our practical aims we limit ourselves with the calculation
of the one-loop expressions for the propagators and vertex 
functions. This is sufficient for reconstruction of the equations 
in the LLA using the renormalization group invariance. 

The unrenormalized fermion Green function is given by the expression
\begin{equation}
G^{-1}(k)=\not\! k-m_0-\Sigma (k),
\end{equation}
where $\Sigma (k)$ is a self-energy operator.
Calculating the latter to the one-loop accuracy with a
principal value (PV) prescription \cite{bas91} for the auxiliary 
pole in the gluon propagator in light-like gauge we get the 
following result
\begin{equation}
G(k)=(1-\Sigma_1)U_2^{-1}(k)\frac{1}{\not\! k-m}U_1(k),
\end{equation}
where 
\begin{eqnarray}
&&U_1 (k) =1-\frac{m}{k_+}\Sigma_2(k)\gamma_+
-\frac{1}{k_+}(\Sigma_2(k)-\Sigma_1)\gamma_+ \!\!\!\not\! k,\\
&&U_2 (k) =1+\frac{m}{k_+}\Sigma_2(k)\gamma_+
+\frac{1}{k_+}(\Sigma_2(k)-\Sigma_1)\not\! k \gamma_+,
\end{eqnarray}
and
\begin{equation}
\Sigma_1 =\frac{\alpha}{4\pi}\ln \Lambda^2,\hspace{0.3cm}
\Sigma_2(k) =\frac{\alpha}{4\pi}\ln \Lambda^2
\int dz' \frac{z}{(z-z')}\Theta_{11}^0 (z',z'-z).
\end{equation}
Here $m$ is a renormalized fermion mass related to the bare  
quantity by the well known relation
\begin{equation}
m_0 =m (1-3\Sigma_1).
\end{equation}
The functions $\Theta^m_{i_1 i_2 ... i_n }$ and some useful relations
between them are written down explicitly in the appendix. The 
renormalization constants are not numbers any more but matrices 
acting on the spinor indices of fermion field operators.

An abelian Ward identity leads to the equality of the renormalization 
constants of the gauge boson wave-function and a charge $Z_3 = 
Z_{\rm g}$, so, that the corresponding logarithmic dependence on the UV 
cut-off cancels in their sum in the evolution equation and, therefore, 
we can neglect the fermion loop insertions into the boson line 
(in the QCD case this is no longer true).

One can easily calculate the vertex function to the same accuracy.
The result is
\begin{equation}
\Gamma_\rho (k_1,k_2)=(1+\Sigma_1)U_1^{-1}(k_1){\cal G}_\rho U_2(k_2),
\end{equation}
where
\begin{equation}
{\cal G}_\rho
=\gamma_\rho
-(\not\! k_1 -m){\cal Q}_\rho(k_1,k_2) \gamma_+
-\gamma_+ {\cal Q}_\rho(k_1,k_2) (\not\! k_2 -m)
\end{equation}
and
\begin{equation}
{\cal Q}_\rho(k_1,k_2)
=\Sigma_3 (k_1) \gamma_- \gamma_+ \gamma_\rho^\perp
+\Sigma_3 (k_2) \gamma_\rho^\perp  \gamma_+ \gamma_-,
\end{equation}
here
\begin{equation}
\Sigma_3 (k_i)
=\frac{\alpha}{8\pi}\ln \Lambda^2
\int dz' \frac{(z_i-z')}{z'}\Theta_{111}^1 (z',z'-z_1,z'-z_2).
\end{equation}
Apart from the graphs we are accounted for there exists an additional  
UV divergence of the virtual Compton scattering amplitude, however,
we do not need its explicit expression for our practical purposes. 
This completes the consideration of virtual corrections which cause 
the logarithmic dependence on the UV momentum cut-off of the 
quantities in question.

\subsection{Sample calculation of the evolution kernels}

As we have noted above the UV divergences also occur in the 
transverse-momentum integrals of partons interacting with a bare cut 
vertex. To extract this dependence properly it is sufficient to 
separate the perturbative loop from correlation function in 
questions. To this end the latter can be represented in the form of 
momentum integral in which the integration
over the fractional energies of the particles attached to the 
vertex is removed
\begin{equation}
\left(
\begin{array}{c}
{\cal I} (\zeta)\\
{\cal M} (\zeta)
\end{array}
\right)
=
\int \frac{d^4k}{(2\pi)^4}
\delta (\zeta - z)
\left( 
\begin{array}{c}
I\\
\frac{1}{\zeta}m\gamma_+
\end{array}
\right)
F(k),
\end{equation}
where
\begin{equation}
F(k)=
\int d^4 xe^{ikx}
\langle 0|
\psi (x)
|h,X\rangle
\langle h,X | \bar\psi (0) |0 \rangle  .
\end{equation}

In the same way we can easily write corresponding expressions for the
three-particle correlation functions.

Let us consider, for definiteness, the fracture function ${\cal I}$.
Simple calculation of the one-loop diagram for the $2 \to 2$
transition in the LLA gives
\begin{eqnarray}
{\cal I}(\zeta)_{\Lambda^2}&=&{\rm g}^2
\int\frac{d^4 k}{(2\pi)^4}F(k)
\int \frac{d^4k''}{(2\pi)^3}\delta (\zeta - z'')
\theta (z'' -z)\frac{\delta ((k'' - k)^2)}{k''^4}\nonumber\\
&&\times\{ -d_{\mu \nu } (k'' - k) \gamma_\mu (\not\! k''+m) 
I (\not\! k'' +m)\gamma_\nu \}\nonumber\\
&&\nonumber\\
&=& - \frac{\alpha}{2\pi^2}\int \frac{d^4k}{(2\pi)^4}
F(k)\int dz'' 
\frac{\delta (z'' - \zeta) \theta (z'' -z )}{(z'' - z)}
\int d^2k''_\perp \int d\alpha'' 
\delta \left(\alpha'' +\frac{k''^2 _\perp}{2(z''-z)}\right)\nonumber\\
&&\times\frac{1}{[2\alpha'' z'' + k''^2_\perp]^2}
\left\{ 
[2\alpha'' z'' + k''^2_\perp] 
- 2m\gamma_+ \left[\alpha'' 
+\frac{[2\alpha'' z'' + k''^2_\perp]}{(z'' - z)} \right]
\right\}\nonumber\\
&&\nonumber\\
&=&-\frac{\alpha}{2\pi}\ln \Lambda^2
\int \frac{dz}{z}\theta (\zeta - z)
\left[
{\cal I}(z)
-{\cal M}(z)\
\left(
1+2\frac{z}{(\zeta - z)}
\right) \label{two}
\right].
\end{eqnarray}
As long as logarithmic contribution appears when $|k_\perp|/ 
|k''_\perp| \ll 1$ and $\alpha / \alpha''\ll 1$ we expand the 
integrand in powers of these ratios keeping the terms that do 
produce the logarithmic divergence.  Similarly one can evaluate 
the transition amplitudes of ${\cal I}$ to the three-particle 
correlation functions ${\cal Z}_j$:
\begin{eqnarray}
{\cal I}(\zeta)_{\Lambda^2}&=&
-\frac{\alpha}{2\pi}\ln \Lambda^2
\int dz dz' \theta (\zeta - z)
{\cal Z}_1 (z', z)
\left[ 
\frac{2}{(\zeta - z)}\label{three}
+
\frac{1}{(z-z')},
\right]\\
{\cal I}(\zeta)_{\Lambda^2}&=&
-\frac{\alpha}{2\pi}\ln \Lambda^2
\int dz dz' \theta (\zeta - z)
{\cal Z}_2 (z, z')
\frac{(\zeta - z)}{(\zeta -z')(z -z')}.
\end{eqnarray}
Due to the non-quasi-partonic \cite{lip85} form of the vertex $I$
there exists an additional contribution to the evolution
equations coming from the contact terms resulting from the 
cancellation of the propagator adjacent to the quark-gluon and
bare cut vertices. As an output the vertex acquires the three-particle
piece
\begin{eqnarray}
{\cal I}(\zeta)_{\Lambda^2}
&=&\int \frac{d^4k}{(2\pi)^4} \frac{d^4k'}{(2\pi)^4}
{\cal Z}_{1\rho}(k',k)
i{\rm g}\Gamma_\rho (k-k', k)i G(k)\delta (\zeta - z) + (c.c.)\nonumber\\
&=& -\frac{\alpha}{2\pi}\ln\Lambda^2
\int dz' {\cal Z}_1(z', \zeta) 
\int \frac{\zeta (\zeta - z'- z'')}{z''}
\Theta^0_{111} (z'' ,z''-\zeta ,z''- \zeta +z').
\end{eqnarray}
As can be seen eqs. (\ref{two}) and (\ref{three}) possess the IR
divergences at $z =\zeta$. They disappear after we account for 
the virtual radiative corrections (renormalization of the field 
operators) discussed in the previous subsection. The net result 
looks like
\begin{equation}
\Gamma^R = (1 - \Sigma_1)U_1\Gamma U^{-1}_2, \hspace{0.5cm}
\Gamma = \left(I,\hspace{0.2cm} \frac{1}{\zeta}m \gamma_+ ,\hspace{0.2cm}
{\rm g} \gamma^\perp_\rho \gamma_+  \right).
\label{renorm}
\end{equation}

Assembling all these contributions we come to the evolution equation 
for ${\cal I}$ given below by eq. (\ref{iev}).

\subsection{Evolution equations.}

Now following the procedure just described it is not difficult to 
construct the closed set of the evolution equations
\begin{eqnarray}
\dot{{\cal M}}(\zeta)
&=&\frac{\alpha}{2\pi}\int \frac{dz}{z}\theta (\zeta - z)
P_{\cal MM}\left(\frac{\zeta}{z}\right)\label{mass}
{\cal M}(z),\\
&&\nonumber\\
\dot{{\cal I}}(\zeta)\label{iev}
&=&\frac{\alpha}{2\pi}\int \frac{dz}{z}\theta (\zeta - z)
\biggl\{
P_{\cal II}\left(\frac{\zeta}{z}\right){\cal I}(z)
+ P_{\cal IM}\left(\frac{\zeta}{z}\right) {\cal M}(z) \nonumber\\
&-&\int dz'\biggl[
P_{{\cal I} {\cal Z}_1}\left(\frac{\zeta}{z},\frac{z'}{\zeta}\right)
{\cal Z}_1(z',z)
-\frac{z(\zeta - z)}{(\zeta - z')(z - z')}
{\cal Z}_2(z,z')
\biggr]
\biggr\},\\
&&\nonumber\\
\dot{{\cal Z}}_1(\zeta',\zeta)&=&
\frac{\alpha}{2\pi}
\biggl\{
\Theta_{11}^0 (\zeta',\zeta'-\zeta)
\left[ \frac{(\zeta - \zeta')}{\zeta}{\cal I}(\zeta) 
- {\cal M}(\zeta) \right]\nonumber\\
&+&
\theta (\zeta')
\left[ \frac{1}{(\zeta - \zeta')}{\cal I}(\zeta - \zeta')
- \frac{1}{\zeta}{\cal M}(\zeta - \zeta') \right] 
\nonumber\\
&+&\int \frac{dz}{z} \theta (\zeta - z)
\biggl[
P_{{\cal Z}_1{\cal Z}_1} 
 \left( \frac{\zeta}{z}, \frac{\zeta'}{\zeta} \right)
{\cal Z}_1 (\zeta', z)
-
\frac{z(\zeta - z)^2}{\zeta \zeta' (z - \zeta + \zeta')}
{\cal Z}_2 (z, \zeta - \zeta')
\biggr]\nonumber\\
&+&\int dz' \biggl[
\Theta_{111}^0 (\zeta',\zeta'-\zeta,\zeta'-\zeta+z')
\frac{(\zeta'-\zeta+z')}{\zeta'}
{\cal Z}_1 (z', \zeta)\nonumber\\
&+&
\theta (\zeta')
\frac{(z' - \zeta)}{\zeta (z' - \zeta + \zeta')}
{\cal Z}_1 (z', \zeta - \zeta')
-
\theta (\zeta - \zeta')
\frac{\zeta'(\zeta - \zeta')}{\zeta (\zeta - z')(\zeta' - z')}
{\cal Z}_2 (\zeta',z')\label{log}
\biggr]
\biggr\},\\
&&\nonumber\\
\dot{{\cal Z}}_2(\zeta,\zeta')&=&
\frac{\alpha}{2\pi}
\biggl\{
\theta (\zeta - \zeta')
\biggl[
\frac{(\zeta' - \zeta)}{\zeta \zeta'}{\cal I}(\zeta ')
-\frac{1}{\zeta}{\cal M}(\zeta')
\biggr]\nonumber\\
&-&\int \frac{dz}{z} \theta (\zeta -z)\frac{z(z-\zeta)}{\zeta^2}
{\cal Z}_1(z-\zeta,z)\nonumber\\
&+&\int dz' 
\biggl[
\theta (\zeta -\zeta')
\frac{(\zeta-\zeta')(\zeta-\zeta'+z')}{\zeta^2 z'}
{\cal Z}_1(z',\zeta')\nonumber\\
&-&\Theta^0_{11}(\zeta',\zeta'-z')
\frac{\zeta'}{(\zeta' - z')}
\left[{\cal Z}_2(\zeta,z')
-{\cal Z}_2(\zeta,\zeta')\right]\nonumber\\
&-&\Theta^0_{11} (\zeta' - \zeta, \zeta' - z')
\frac{(\zeta' - \zeta)}{(\zeta' - z')}
\left[
{\cal Z}_2(\zeta,z')
-{\cal Z}_2(\zeta, \zeta')
\right]
\biggr]
+\frac{3}{2}{\cal Z}_2(\zeta, \zeta')
\biggr\}.
\end{eqnarray}
where the dot denotes the derivative with respect to the UV cutoff
$\hspace{0.1cm}\dot{}=\Lambda^2{\partial}/{\partial\Lambda^2}$ and
splitting functions are given by the following equations
\begin{eqnarray}
P_{\cal MM}(z)
&=&-\left[\frac{2}{z(1-z)}\right]_+ + \frac{1}{z} + 1,\\
P_{\cal II}(z)
&=& -1 + \frac{1}{2}\delta (z-1),\\
P_{\cal IM }(z)
&=&-\left[\frac{2}{z(1-z)}\right]_+ + \frac{2}{z} + 1, \\
P_{{\cal I} {\cal Z}_1}(z,y)
&=&-\left[\frac{2}{z(1-z)}\right]_+ + \frac{2}{z}
+ \frac{1}{1-yz} - \delta (z-1) \frac{1}{y}\ln (1-y),\\
P_{{\cal Z}_1 {\cal Z}_1}(z,y)
&=&-\left[\frac{2}{z(1-z)}\right]_+ + \frac{2}{z}
+\frac{y}{1-yz}
+ \delta (z-1) \left[ \frac{3}{2} - \ln (1-y) \right].
\end{eqnarray}

Now it is an easy task to verify the fulfilment of the equation of
motion (\ref{eqmot}) for the correlation functions as a consistency
check of our calculations. By exploiting this relation we exclude 
${\cal I}$ from the above set of functions and reduce the system 
to the basis of independent gauge invariant quantities 
$\{{\cal M}$, ${\cal R}_j\}$. 

An important note is in order now. As distinguished from the 
DIS case the above eq. (\ref{log}) has the logarithmic dependence 
on the ratio of the parton momentum fractions. The consequence 
of its presence is obvious. Taking into account the restrictions 
imposed by eq. (\ref{pos}) we can define the moments of the correlation
functions in the following way
\begin{eqnarray}
&&{\cal M}_n =\int_{1}^{\infty}
\frac{d\zeta}{\zeta^n}{\cal M}(\zeta),\\
&&{\cal R}^m_n =\int_{1}^{\infty}\frac{d\zeta}{\zeta^n}
\int_{0}^{\zeta} d \zeta' \zeta'^m {\cal R}(\zeta',\zeta).
\end{eqnarray}
We find for two-particle cut vertex
\begin{eqnarray}
&&\dot{\cal M}_n
=\frac{\alpha}{2\pi}
\left[
-\psi (n+1) - \psi (n-1)-2\gamma_E
\right]{\cal M}_n.
\end{eqnarray}
where $\gamma_E$ is a Euler-Masceroni constant and
$\psi(n)=\Gamma'(n)/\Gamma(n)$. And we see the universality of the
evolution kernels for the time- and space-like quasi-partonic 
two-particles cut vertices, {\it i.e.} the Gribov-Lipatov reciprocity 
relation is fulfilled \cite{grlip72_2}. However, 
it is impossible to write down the finite system of equations for 
any moment of the three-parton correlation functions
as the logarithm of the ratio of the parton momentum fractions 
in the evolution kernels leads to the infinite series of the moments 
as distinguished from the deep inelastic scattering where the rang of 
the anomalous dimension matrix was finite and increases as the moment 
of the correlation function increases. Therefore, it is not possible 
to solve the system of equations successively in terms of moments as 
well as we do not succeed in solving it analytically in a general 
form. However, in the next section when dealing with the QCD evolution 
we will find that the system can be reduced to the single equation 
in the limit of infinite number of colours and there is believe that 
its solution can be found analytically.

\section{Non-abelian evolution.}

For the non-abelian gauge theory the equality of the renormalization
constants $Z_{\rm g} = Z_3$ no longer holds, so, we should 
account for the renormalization of the gluon wave-function as well as 
for the renormalization of charge explicitly.
For this purposes to complete the renormalization program outlined 
in the preceding section we evaluate the gluon propagator to the 
same accuracy. The result can be written in the compact form
\begin{equation}
D_{\mu\nu}(k) = \left( 1+\Pi^{tr}(k) \right)
U_{\mu\rho}(k)\frac{d_{\rho\sigma}(k)}{k^2+i0}
U_{\sigma\nu}(k),
\end{equation}
where
\begin{equation}
U_{\mu\nu}(k)=g_{\mu\nu}-\frac{1}{2}\Pi^{add}(k)
\frac{k_\mu n_\nu + k_\nu n_\mu}{k_+}
\end{equation}
and
\begin{eqnarray}
&&\Pi^{tr}(k)=2 \frac{\alpha}{4\pi} \ln \Lambda^2
\left\{
C_A\int dz \frac{[z^2-z\zeta +\zeta^2]^2}{z(z-\zeta)\zeta^2}
\Theta_{11}^0 (z,z-\zeta)
-\frac{N_f}{3}
\right\},\nonumber\\
&&\Pi^{add}(k)= \frac{\alpha}{4\pi} \ln \Lambda^2
C_A\int dz
\frac{[5z\zeta^2(z-\zeta) + 6z^2(z-\zeta)^2 + 2\zeta^4]}{z(z-\zeta)\zeta^2}
\Theta_{11}^0 (z,z-\zeta)
\end{eqnarray}
are the transverse and longitudinal pieces of polarization operator.\
While the renormalized charge is given by the well known ``asymptotic 
freedom'' formula
\begin{equation}
{\rm g}_0
={\rm g}\left[
1 + \frac{\alpha}{4\pi}\ln \Lambda^2
\left( 
\frac{N_f}{3} - \frac{11}{6}C_A
\right)
\right].
\end{equation}

The second addendum arises from the diagrams with
triple-boson interaction vertex. Gathering these contributions 
together with equations obtained in the previous section (with colour 
group factors accounted for properly), we come to the final result
\begin{eqnarray}
\dot{\cal Z}_1&\!\!\!\!\!(&\!\!\!\!\!\zeta',\zeta)=
\frac{\alpha}{2\pi}
\Biggl\{
C_F
\left[
\theta (\zeta') \frac{\zeta'}{\zeta (\zeta - \zeta')}
{\cal M}(\zeta -\zeta')
- \Theta^0_{11} (\zeta', \zeta' - \zeta)\frac{\zeta'}{\zeta}
{\cal M}(\zeta)
\right]\nonumber\\
&\!\!\!\!\!+\!\!\!\!\!&\int \frac{dz}{z}\theta (\zeta - z)
\Biggl[
P_{{\cal Z}_1{\cal Z}_1}\left( \frac{\zeta}{z}, \frac{\zeta'}{\zeta} \right)
{\cal Z}_1 (\zeta' ,z)
- \left( C_F -\frac{C_A}{2} \right)
\frac{z(\zeta - z)^2}{\zeta \zeta' (z - \zeta' + \zeta)}
{\cal Z}_2(z,\zeta - \zeta')\nonumber\\
&\!\!\!\!\!+\!\!\!\!\!&\frac{C_A}{2}
\Biggl(
-2z \frac{\partial}{\partial\zeta'}\int_{0}^{1}dv
{\cal Z}_1(\zeta' - v(\zeta - z), z)
+\left(\frac{z}{(z-\zeta+\zeta')} -\frac{z(z+\zeta')}{\zeta \zeta'} \right)
{\cal Z}_1(z- \zeta+\zeta',z)
\Biggr)
\Biggr]\nonumber\\
&\!\!\!\!\!+\!\!\!\!\!&\int dz'
\Biggl[
-C_F
\left(
\Theta^0_{11} (\zeta',\zeta'-\zeta)\frac{(\zeta - \zeta')}{\zeta}
{\cal Z}_1(z',\zeta)
+\theta (\zeta')\frac{1}{(\zeta- \zeta')}
{\cal Z}_1(z',\zeta-\zeta')
\right)\nonumber\\
&\!\!\!\!\!+\!\!\!\!\!& \left(C_F - \frac{C_A}{2} \right)
\Biggl(
\Theta^0_{111}(\zeta', \zeta' - \zeta, \zeta' - \zeta +z')
\frac{(\zeta' - \zeta +z')}{\zeta'}
{\cal Z}_1(z',\zeta)\nonumber\\
&\!\!\!\!\!+\!\!\!\!\!& \theta (\zeta')
\frac{(z' -\zeta)}{\zeta (z' - \zeta + \zeta')}
{\cal Z}_1(z', \zeta- \zeta')
-\theta (\zeta - \zeta')
\frac{\zeta' (\zeta - \zeta')}{\zeta (\zeta -z')(\zeta' -z')}
{\cal Z}_2(\zeta', z')
\Biggr)\nonumber\\
&\!\!\!\!\!+\!\!\!\!\!&\frac{C_A}{2}
\Biggl(
\Theta^0_{111}(\zeta', \zeta' -\zeta, \zeta' -z')
\frac{z'(z' +\zeta' - \zeta)}{\zeta' (\zeta' - z')}
{\cal Z}_1(z',\zeta)\nonumber\\
&\!\!\!\!\!+\!\!\!\!\!&\Theta^0_{11}(\zeta', \zeta' - \zeta)
\frac{(\zeta - \zeta')(\zeta' + z')}{(\zeta' - z')(\zeta - z')}
{\cal Z}_1(z',\zeta)
+\Theta^0_{11}(\zeta',\zeta' - z')
\frac{(\zeta'+\zeta)}{(\zeta -z')}{\cal Z}_1(z',\zeta)\nonumber\\
&\!\!\!\!\!+\!\!\!\!\!&\theta (\zeta')
\left( \frac{1}{(z'-\zeta')} - \frac{(\zeta - \zeta' +z')}{\zeta z'} \right)
{\cal Z}_1(z' -\zeta' , \zeta - \zeta')\nonumber\\
&\!\!\!\!\!-\!\!\!\!\!&2\Theta^0_{11}(\zeta', \zeta' - z')
\frac{\zeta'}{(\zeta' - z')}
[{\cal Z}_1(z',\zeta)-{\cal Z}_1(\zeta' , \zeta)]
\Biggr)
\Biggr]
\Biggr\},\\
&&\nonumber\\
\dot{\cal Z}_2&\!\!\!\!\!(&\!\!\!\!\!\zeta, \zeta')=
\frac{\alpha}{2\pi}
\Biggl\{
-C_F \theta(\zeta - \zeta')\frac{1}{\zeta'}{\cal M}(\zeta)\nonumber\\
&\!\!\!\!\!+\!\!\!\!\!&
\int \frac{dz}{z}\theta (\zeta - z)
\Biggl[
P_{{\cal Z}_2{\cal Z}_2}
\left( \frac{\zeta}{z},\frac{\zeta'}{\zeta} \right)
{\cal Z}_2(\zeta ,z)
-\left(C_F -\frac{C_A}{2} \right)
\frac{z(z-\zeta)}{\zeta^2}{\cal Z}_1(z -\zeta,z)\nonumber\\
&\!\!\!\!\!+\!\!\!\!\!&
\frac{C_A}{2}
\Biggl(
-2z\frac{\partial}{\partial\zeta'}\int_0^1 dv
{\cal Z}_2 (z,\zeta' -v(\zeta -z))
+\left(
\frac{z}{(z-\zeta +\zeta')}-\frac{z(z+\zeta)}{\zeta^2}
\right)
{\cal Z}_2(z, z-\zeta+\zeta')
\Biggr)
\Biggr]\nonumber\\
&\!\!\!\!\!+\!\!\!\!\!&
\int dz'
\Biggl[
C_F\theta (\zeta - \zeta')\frac{\zeta - \zeta'}{\zeta \zeta'}
{\cal Z}_1(z', \zeta')
\nonumber\\
&\!\!\!\!\!+\!\!\!\!\!&
\left(C_F - \frac{C_A}{2} \right)
\Biggl(
\theta (\zeta -\zeta')
\frac{(\zeta - \zeta')(\zeta - \zeta'+z')}{\zeta^2 z'}
{\cal Z}_1(z', \zeta')\nonumber\\
&\!\!\!\!\!-\!\!\!\!\!&
\Theta^0_{11} (\zeta' , \zeta'-z')\frac{\zeta'}{(\zeta' - z')}
[{\cal Z}_2(\zeta ,z')-{\cal Z}_2(\zeta , \zeta')]
\nonumber\\
&\!\!\!\!\!-\!\!\!\!\!&
\Theta^0_{11}(\zeta' - \zeta, \zeta' - z')
\frac{(\zeta' - \zeta)}{(\zeta' -z')}
[{\cal Z}_2(\zeta, z')-{\cal Z}_2(\zeta ,\zeta')]
\Biggr)\nonumber\\
&\!\!\!\!\!-\!\!\!\!\!&
\frac{C_A}{2}
\theta (\zeta -\zeta')
\frac{(\zeta - \zeta')}{(\zeta - z')}
\left(
\frac{1}{(\zeta' - z')}
+\frac{z'}{\zeta^2}
\right)
{\cal Z}_1(z',\zeta')
\Biggr]
\Biggr\},
\end{eqnarray}

where

\begin{eqnarray}
P_{{\cal Z}_1{\cal Z}_1}(z,y)
&=&C_F \left\{ 
-\left[\frac{2}{z(1-z)} \right]_+ +\frac{2}{z}
+\delta(z-1)\left[ \frac{3}{2} -\ln (1-y)\right]
\right\}\nonumber\\
&+&\left( C_F -\frac{C_A}{2}\right)
\frac{y}{1-yz},\nonumber\\
P_{{\cal Z}_2{\cal Z}_2}(z,y)
&=&\frac{C_A}{2}
\left\{
-\left[
\frac{4}{z(1-z)}
\right]_+
+\frac{4}{z}
+\frac{1}{1-yz}
-\frac{1+z}{z^2}
\right\}\nonumber\\
&+&\delta (z-1)
\left[
\frac{3}{2}C_F
-\frac{C_A}{2}
(\ln y + \ln (1-y ))
\right].
\end {eqnarray}
These equations should be supplemented by the equation for the mass cut
vertex ${\cal M}$ which differs from its abelian analogue 
(\ref{mass}) only by the group factor $C_F$. 

One can easily observe the significant simplification of the above 
evolution equations in the limit $N_c \to \infty$
since ${\cal Z}_2$ decouples from the evolution equation for
${\cal Z}_1$. Therefore, discarding the quark mass cut vertex 
we obtain homogeneous equation which governs the $Q^2$-dependence
of the three-parton correlation function ${\cal Z}_1$.
Similar phenomenon has been found in the evolution 
equations for chiral-even and -odd distribution functions in DIS 
\cite{bbkt96,ali91} where it 
has been observed that for multicolour QCD the momentum fraction 
carried by gluon in matrix element of quark-gluon operator varies 
only among the quarks ones and does not exceed the latter. Owing to 
this feature the solution of approximate equations becomes possible. 
In the forthcoming paper we will address to this issue in greater 
details.

\section{Conclusion.}

We have developed the effective technique for construction of the 
evolution equations for the twist-3 cut vertices. The physically
transparent picture which appears in the light-cone gauge 
(an essential ingredient of our method) makes the calculations simple.
Using this technique we have constructed the basis of chiral-odd
non-polarized cut vertices closed under renormalization group
evolution. The identity provided by the equation of motion for
field operators makes the construction of the basis of independent
operators trivial.

The most striking difference from the DIS evolution is the
appearance of the logarithmic dependence on the parton
momentum fractions in the evolution kernels which makes the
successive solution of equations in terms of moments impossible.
The nonlocality of the cut vertex in coordinate space is essential 
since even if we start from the local cut vertex it will smeared along 
the light cone upon the renormalization.

However, an important point is that there is decoupling of three
parton correlation function ${\cal Z}_1$ from ${\cal Z}_2$.
The situation has the closer similarity with DIS where in the
limit $N_c \to \infty$ there was a very important simplification
as the evolution kernels have been vanishing for contributions
with interchanged order of partons on the light cone, {\it i.e.} the 
gluon momentum fraction ranges between those of the quarks only.
This property allowed the author of refs. \cite{bbkt96,ali91}
to find the solution of simplified equations exactly on the nonlocal 
form. In our case the decoupling of ${\cal Z}_1$ may have the same 
consequences. This work is in progress now and the results will be
published elsewhere.

\hspace{0.5cm}

We are grateful to L.N. Lipatov for helpful discussions and 
encouraging. One of the authors (A.B.) would like to thank CERN
Theory Division for hospitality extended to him during his
visit where this work was finished. He benefited from useful 
discussion with V.M. Braun, D. M\"uller and M.A. Shifman. This 
work was supported by Russian Foundation for Fundamental Research, 
grant N 96-02-17631.

\appendix

\section{Appendix}

In this appendix we present the necessary properties of the
$\Theta$-functions
\begin{equation}
\Theta^{m}_{i_1 i_2 ... i_n}
(z_1,z_2,...,z_n)=\int_{-\infty}^{\infty}\frac{d\alpha}{2\pi i}
\alpha^m \prod_{k=1}^{n}\left(\alpha z_k -1 +i0 \right)^{-i_k}.
\end{equation}
For our discussion it is enough to have an explicit form of the
function
\begin{equation}
\Theta^0_{11} (z_1,z_2)
=\frac{\theta(z_1)\theta(-z_2) -\theta(z_2)\theta(-z_1)}{z_1-z_2},
\end{equation}
since the others can be expressed in its term via relations
\begin{eqnarray}
&&\Theta^0_{21} (z_1,z_2)=\frac{z_2}{z_1-z_2}\Theta^0_{11} (z_1,z_2),\\
&&\Theta^0_{111} (z_1,z_2,z_3)
=\frac{z_2}{z_1-z_2}\Theta^0_{11} (z_2,z_3)
-\frac{z_1}{z_1-z_2}\Theta^0_{11} (z_1,z_3),\\
&&\Theta^1_{111} (z_1,z_2,z_3)
=\frac{1}{z_1-z_2}\Theta^0_{11} (z_2,z_3)
-\frac{1}{z_1-z_2}\Theta^0_{11} (z_1,z_3).
\end{eqnarray}
In the main text we have used the identity
\begin{equation}
{\rm PV}\int dz \frac{\zeta}{(\zeta - z)}
\left[
\Theta^0_{11} (z,z-\zeta)
+\Theta^0_{11}(\zeta,\zeta-z)
\right]=0.
\end{equation}


\begin{thebibliography}{99}
\bibitem{wil69}
K. Wilson, Phys. Rev. 179 (1969) 1499;\\
R. Brandt, G. Preparata, Nucl. Phys. B 27 (1971) 541.
\bibitem{mue89}
A.H. Mueller, ed., {\it Perturbative Quantum Cromodynamics}
(World Scientific, Singapore, 1989).
\bibitem{lip74}
L.N. Lipatov, Sov. J. Nucl. Phys. 20 (1974) 94;\\
A.P. Bukhvostov, L.N. Lipatov, N.P. Popov, Sov. J. Nucl. Phys.
20 (1974) 287.
\bibitem{alt77}
G. Altarelli, G. Parisi, Nucl. Phys. B 126 (1977) 298.
\bibitem{muel78}
A.H. Mueller, Phys. Rev. D 18 (1978) 3705.
\bibitem{muel81}
A.H. Mueller, Phys. Rept. 73 (1981) 237.
\bibitem{bal91}
I.I. Balitsky, V.M. Braun, Nucl. Phys. B 361 (1991) 93.
\bibitem{cfp80}
G. Curci, W. Furmanski, R. Petronzio, Nucl. Phys. B 175 (1980) 27.
\bibitem{koi95}
Y. Koike, K. Tanaka, Phys. Rev. D 51 (1995) 6125;\\
Y. Koike, N. Nishiyama, {\it $Q^2$-evolution of the chiral odd
twist-3 distribution $e(x,Q^2)$}, hep-ph/9609207.
\bibitem{bbkt96}
I. Balitsky, V. Braun, Y. Koike, K. Tanaka,
Phys. Rev. Lett. 77 (1996) 3078.
\bibitem{ji94}
X. Ji, Phys. Rev. D 49 (1994) 114.
\bibitem{jaf91}
R.L. Jaffe, X. Ji, Phys. Rev. Lett. 67 (1991) 552;\\
R.L. Jaffe, X. Ji, Nucl. Phys. B 357 (1992) 527.
\bibitem{lip84}
A.P. Bukhvostov, E.A. Kuraev, L.N. Lipatov,
Sov. J. Nucl. Phys. 38 (1983) 263; {\it ibid.} 39 (1984) 121.
\bibitem{lip85}
A.P. Bukhvostov, E.A. Kuraev, L.N. Lipatov,
JETP Lett. 37 (1983) 482; Sov. Phys. JETP 60 (1984) 22;\\
A.P. Bukhvostov, G.V. Frolov, L.N. Lipatov, E.A. Kuraev,
Nucl. Phys. B 258 (1985) 601.
\bibitem{jaf83}
R.L. Jaffe, Nucl. Phys. B 229 (1983) 205.
\bibitem{grlip72_1}
V.N. Gribov, L.N. Lipatov, Sov. J. Nucl. Phys. 15 (1972) 438.
\bibitem{grlip72_2}
V.N. Gribov, L.N. Lipatov, Sov. J. Nucl. Phys. 15 (1972) 675.
\bibitem{bas91}
For a review, see A. Bassetto, G. Nardelli, R. Soldati, 
{\it Yang-Mills Theories in the Algebraic non Covariant Gauges} 
(World Scientific, Singapore, 1991).
\bibitem{ali91}
I.I. Balitsky, V.M. Braun, Nucl. Phys. B 361 (1988/89) 541;\\
A. Ali, V.M. Braun, G. Hiller, Phys. Lett. B 266 (1991) 117.
\end{thebibliography}
\end{document}